\documentclass[10pt,twocolumn]{article}
\usepackage{hyperref}
\usepackage{graphicx}
\usepackage{siunitx}
\usepackage{balance}
\usepackage{color, colortbl}

\definecolor{LightCyan}{rgb}{0.88,1,1}
\usepackage[legalpaper, textheight=29cm, textwidth=17cm]{geometry}
\title{Super-resolution imaging within reach}
\author{Stephane Perrin, Keshia Badu, Paul Montgomery, and Sylvain Lecler \\
  ICube laboratory, University of Strasbourg - CNRS, FR-67412 ILLKIRCH \\
  \emph{Corresponding author: }{\tt \href{mailto:stephane.perrin@unistra.fr}{stephane.perrin@unistra.fr}}}
\date{\today}
\begin{document}
\maketitle
\noindent \textbf{Although several optical techniques have been recently developed in order to overcome the resolution limit in microscopy, the imaging of sub-wavelength features is still a real challenge. In practise, super-resolution techniques remain difficult to build or are photo-toxic for the biological samples. However, microsphere-assisted microscopy has recently made super-resolution imaging accessible to scientists (\emph{e.g.} optical metrologists, engineers and biologists). This paper presents an easy-to-implement optical setup to perform full-field and contactless super-resolution measurements of nanostructured media or biological elements. For this purpose, a classical microscope was enhanced by introducing a transparent microsphere. We show that this rather simple approach makes it possible to achieve a lateral resolution of 200~nm in air, \emph{i.e.} the visualization of feature sizes of 100~nm.} 
\par \textbf{Introduction}~~
Considerable progress has been made in optical microscopy in order to overcome the resolution limit imposed by the diffraction of light \cite{SPerrin18}. Several imaging techniques have been proposed in order to visualize beyond the ultimate physical barrier of $\lambda / 2$ \cite{Editorial09}. Among them, we can list confocal microscopy \cite{HGoldmann40} as well as its enhancements (\emph{e.g.}, 4Pi microscopy \cite{SWHell94a}, stimulated-emission-depletion fluorescence microscopy (STED) \cite{SWHell94b}, and photonic-jet-based scanning microscopy \cite{ZChen04}), scanning near-field optical microscopy (SNOM) \cite{EHSynge28,EAAsh72}, metamaterial-based superlenses \cite{VGVeselago68,JBPendry00}, and structured illumination microscopy (SIM) \cite{WLukosz63,BBailey93}. In addition, the Nobel Prize for Chemistry rewarded advances in optical nanoscopy in 2014 \cite{Press14}. However, these super-resolution techniques appear to be complex to set up, need the use of fluorescent labels, are photo-toxic for the sample, or require long-acquisition times. 
\par In 2011, a new imaging technique, microsphere-assisted microscopy, was reported \cite{ZWang11}, experimentally demonstrating the visualization of object features having 50 nm of size in air (in reality, the lateral resolution is 100~nm according to the Abbe criterion \cite{MDuocastella17}). Immersion microsphere-assisted microscopy was suggested shortly after \cite{Darafsheh12}. The physical explanation behind the super-resolution phenomenon is still currently not fully explained. Moreover, the performance of microsphere-assisted microscopy depends on geometrical (\emph{e.g.}, microbead diameter) as well as optical (\emph{e.g.}, wavelength of light or refractive index) parameters. However, the scientific community highlights the probable conversion of the near-field information to propagating waves \cite{Aryeh16}, allowing the lateral resolution to be superior to that attainable by confocal microscopy and by the solid immersion lens \cite{ADarafsheh14}, and to be similar to that achieved from SIM and superlenses. In addition, the influence of the coherence of light on the imaging performance has been studied \cite{Maslov17}, as has been the image formation process, by considering the photonic jet as the focal point of the microsphere \cite{SLecler18}. Furthermore, microsphere-assisted microscopy has been combined with dark-field imaging for the observation of quasi-transparent samples \cite{SPerrin18a} and with interference microscopy for high-resolution topography reconstruction \cite{SPerrin17,Wang16}.
\par This paper features the simple integration of an optical nanoscope using the principle of microsphere-assisted microscopy, \emph{i.e.} by requiring only a light source, a microscope objective with a microlens, and a camera. The design and the optical assembly for achieving a lateral resolution of 200~nm in air are presented. We show that this simple approach can be used to visualize nanostructured surfaces and intra-cellular elements.\\
\begin{figure}[!b]
    \centering
    \includegraphics[scale=1.0]{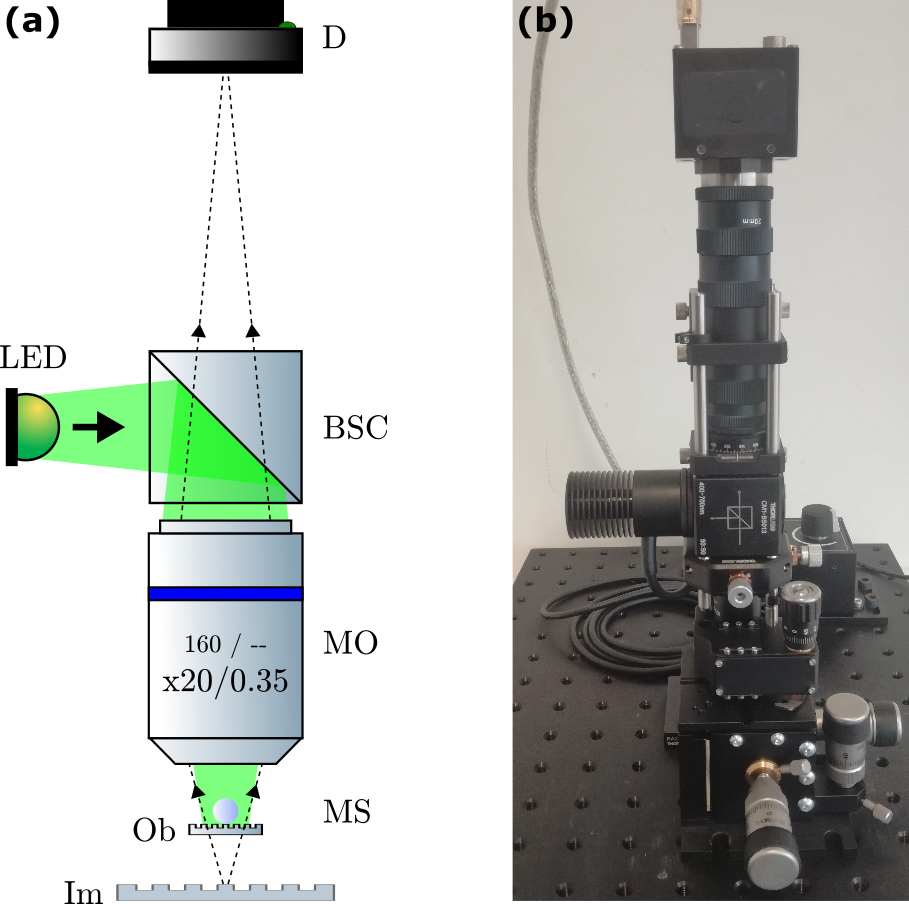}
    \caption{\textbf{Compact optical nanoscope.} \textbf{(a)} Layout of the imaging system. The incident beam from the LED source ($\lambda_{0}$~=~530~nm, $\Delta\lambda$~=~30~nm) is reflected by a beam-splitter cube (BSC) to illuminate the object (Ob) through the microscope objective (MO) and the microlens (MS). Then, the MO collects and directs the virtual image (Im) of the object on the detector (D). \textbf{(b)} Photograph of the imaging system.}
    \label{fig:fig1}
\end{figure}
\par \textbf{Experimental method}~~
The design of the super-resolution optical system slightly differs from an optical microscope only by the introduction of a transparent microsphere. Microsphere-assisted microscopy can thus be seen as an enhancement of white-light microscopy. In this work, the compact nanoscope consists of an illumination part with a light-emitting diode (M530L3, Thorlabs). The wide-field beam is directed by a 50-50 beam-splitter cube onto a microscope objective ($\times$20, NA~=~0.35, S\&H) and then passes through a soda-lime-glass microsphere (SLGMS, Cospheric) in order to illuminate the object. The microsphere is held by a glass plate having a thickness of 150~\si\micro m and axially positioned using a translation mount (SM1Z, Thorlabs) in order to perform contactless measurements. The microsphere must be at a distance of less than one wavelength from the object. The microsphere can also be placed against the object to perform super-resolution imaging. Finally, the microscope objective collects the virtual image generated by the microsphere and propagates the super-resolution information onto a camera (ec1380, Prosilica). The camera is placed at a distance of 160~mm from the microscope objective (the objective lens is not infinity-corrected, and thus does not require a tube lens to form the image on the camera). A graphical program (LabVIEW, National Instruments) has been developed to record the frames as well as to retrieve the intensity profiles.
\begin{figure}[!b]
    \centering
    \includegraphics[scale=0.98]{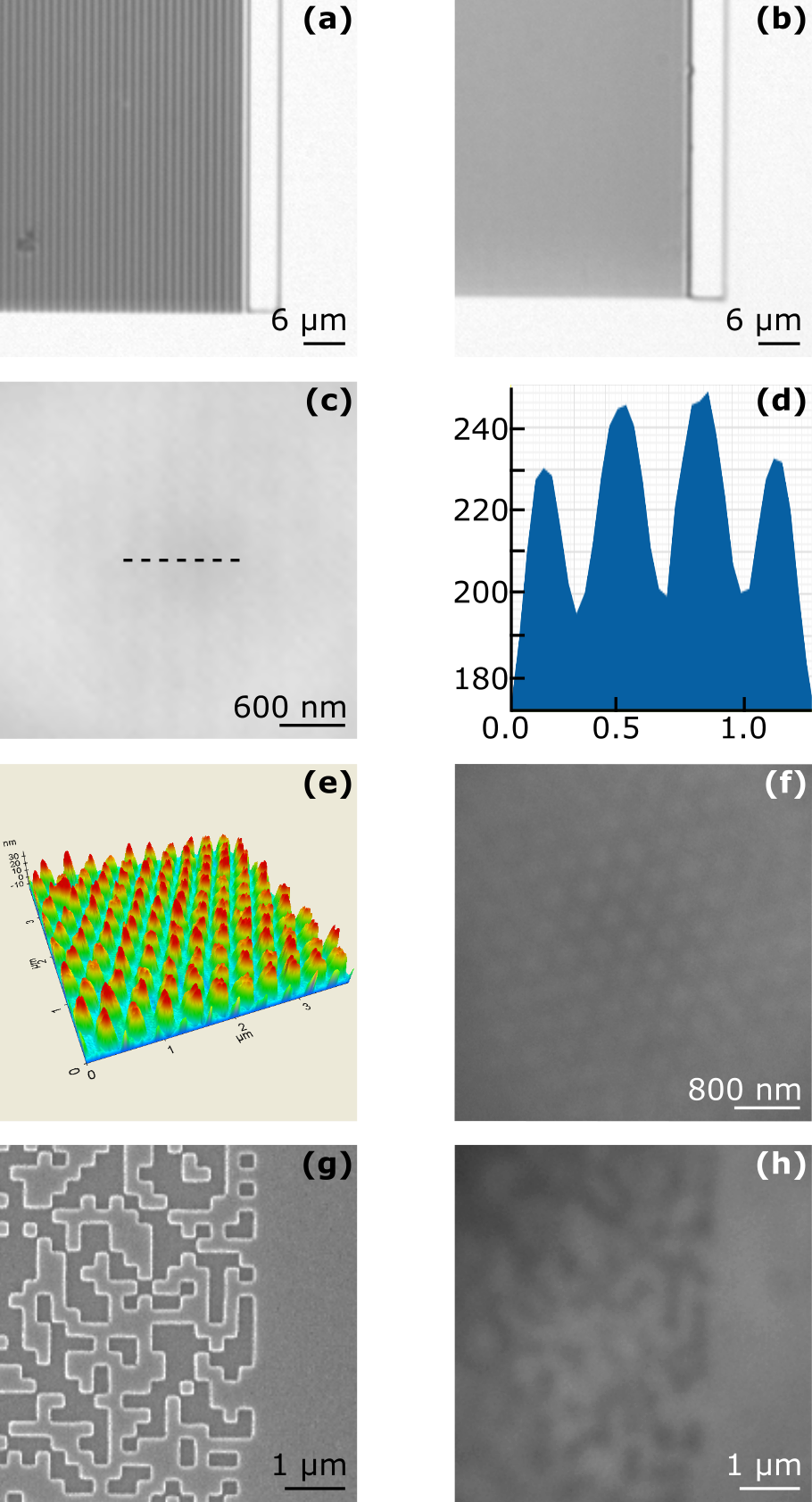}
    \caption{\textbf{Super-resolution direct imaging in air.} \textbf{(a)} 1200-nm-period grating using the $\times$20 objective alone. 300-nm-period grating using the objective \textbf{(b)} alone and \textbf{(c)} combined with the microsphere. \textbf{(d)} Intensity distribution along the black-dashed line in (c). 200-nm nanodots using \textbf{(e)} AFM and \textbf{(f)} the compact nanoscope. Matrix of 250-nm squares using \textbf{(g)} SEM and \textbf{(h)} the optical nanoscope.}
    \label{fig:fig2}
\end{figure}
\par \textbf{Results}~~
Without a microsphere, \emph{i.e.} using the microscope objective alone, the resolving power of the imaging system equals is 1~\si\micro m. Figures \ref{fig:fig2}(a) and \ref{fig:fig2}(b) show the ability of the objective lens to record direct images of a standard silicon grating having a period of 1.2~\si\micro m (features are resolved) and 0.3~\si\micro m (features are not resolved), respectively. Introducing a 25-\si\micro m-diameter microsphere between the microscope objective and the standard, at a few hundred nanometers from the surface of the standard, makes it possible to perform super-resolution and thus, to retrieve the 300-nm-period features (Fig.~\ref{fig:fig2}(c)) with an imaging contrast of around 10\% (Fig.~\ref{fig:fig2}(d)), despite the low contrast of the object. Afterwards, the standard grating was replaced by oval-shaped nanodots of Ag covered by a SiON layer (Fig.~\ref{fig:fig2}(e)). The periodical nanostructures having a size of 200~nm separated by 200~nm, are resolved by the nanoscope (Fig.~\ref{fig:fig2}(f)). A second type of sample was measured, \emph{i.e.} a matrix of random squares with a side of 250~nm (Fig.~\ref{fig:fig2}(g)), in order to validate the super-resolution imaging (Fig.~\ref{fig:fig2}(h)).
\par
\begin{figure}[!h]
    \centering
    \includegraphics[scale=1.0]{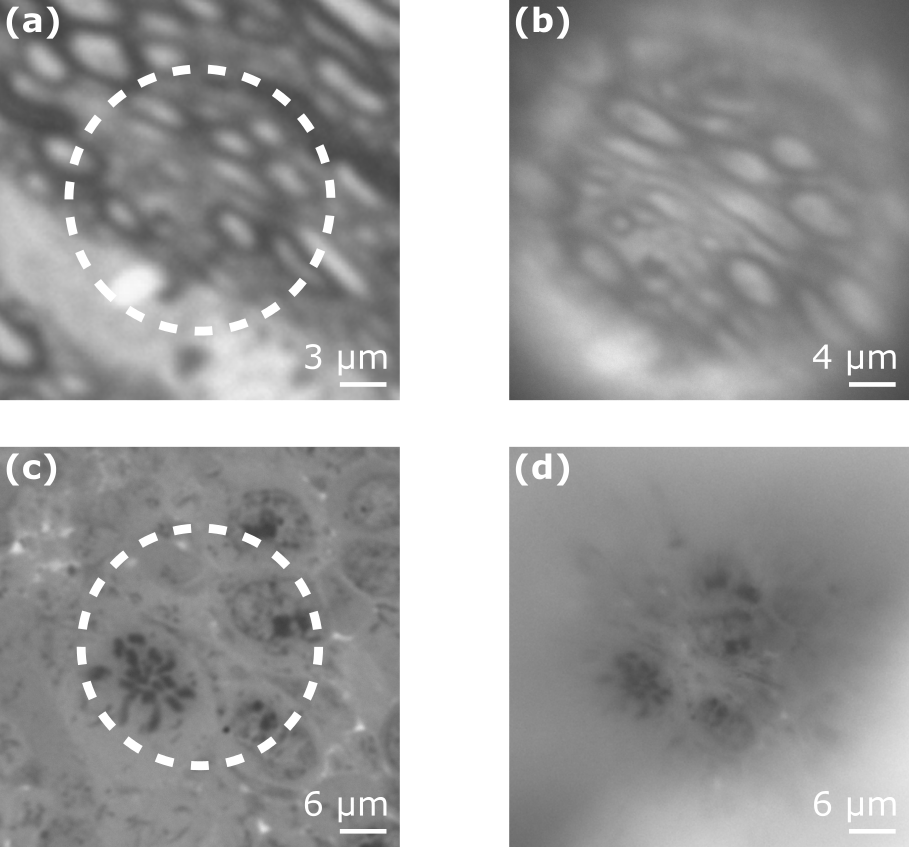}
    \caption{\textbf{Super-resolution direct imaging in air of biological samples.} Myelinated fibers of mouse brain using \textbf{(a)} the $\times$20 objective alone and \textbf{(b)} combined with the microsphere. Mouse embryo cells using \textbf{(c)} a $\times$100 objective alone and \textbf{(d)} the $\times$20 objective combined with a 130-\si\micro m-diameter microsphere. The white-dashed circles represent the lateral fields of view of the microspheres.}
    \label{fig:fig3}
\end{figure}

 The use of microsphere-assisted microscopy appears to be very promising not only for contactless nanometrology, but also for cellular (or even intracellular) imaging. Figure~\ref{fig:fig3}(a) and \ref{fig:fig3}(b) show the direct images of myelinated brain fibres through the optical microscope alone and through the optical nanoscope, respectively. 
The increase in lateral resolution is remarkable and allows the visualization of cells features with a high accuracy. Figure~\ref{fig:fig3}(c) is the direct image of mouse embryo cells through an objective having a higher magnification power ($\times$100, NA~=~0.9, Leica). This was performed in order to compare the results with those in Fig.~\ref{fig:fig3}(d) which represents the image recorded through the $\times$20 objective and a 130-\si\micro m-diameter microsphere. These two measurements show that a ``large'' microsphere (\emph{e.g.}, larger than 60~\si\micro m) provides a similar resolving power as when a high-magnification objective. Nevertheless, the lateral field of view appears to be reduced. A contrast agent (here, uranyl acetate and lead citrate) was required in order to visualize the features of the biological samples (Fig.~\ref{fig:fig3}) with a high imaging contrast.\\
\par \textbf{Conclusion}~~
This paper aims to help scientists in developing an easy-to-implement optical nanoscope based on microsphere-assisted microscopy. The capacity of this super-resolution imaging technique is demonstrated through the visualization of nanostructures and biological elements. A lateral resolution of 200~nm in air is reached. Furthermore, microsphere-assisted microscopy allows the acquisitions to be full-field, \emph{i.e.} no lateral scanning is required, and contactless, \emph{i.e.} not destructive for the samples. Obviously, the resolving power can be improved by performing immersion measurements [\href{https://doi.org/10.1063/1.4757600}{A. Darafsheh \emph{et al.}, Appl. Phys. Lett. \textbf{101} 141128 (2012)}] or nanoplasmonic structures [\href{https://doi.org/10.1109/NAECON.2017.8268767}{A. Brettin \emph{et al.}, Proc. IEEE NAECON (2017)}]. Moreover, a matrix of microspheres can further increase the lateral field of view of the super-resolution system.
\par Table~\ref{tab1} summarizes the components which were used to build the optical nanoscope (on the date of publication \today.)
\begin{table}[ht]
\centering
\begin{tabular}{|r|c|c|}
\hline
\textbf{Items} & \textbf{Reference} & \textbf{Price} \\ 
\hline
Light source & M530L3, & $\sim$~EUR~255 \\ 
& Thorlabs & \\
\hline
Cube & CCM1-BS013, & $\sim$~EUR~255 \\ 
& Thorlabs & \\
\hline
Objective & S\&H, Carl Zeiss & PNA \\ 
\hline
Microsphere & S-SLGMS-2.5, & $\sim$~USD~170 \\ 
& Cospheric & \\
\hline
Camera & ec1380, Prosilica & PNA \\ 
\hline
\end{tabular}
\caption{Summary of the optical components required to build the nanoscope. PNA, price not available.}
\label{tab1} 
\end{table}
\par \textbf{Acknowledgments}~~
The authors thank Giorgio Quaranta (CSEM - EPFL, Switzerland) for providing the QR-code-like structures, Thomas Fix (ICube, France) for the Ag nanodots, and the members from the Imaging and Microscopy Platform (IGBMC, France) for the biological samples. Help from the C$^3$-Fab Platform (ICube, France) is also acknowledged. This work has been funded by SATT Conectus Alsace and supported by University of Strasbourg.
\balance

\end{document}